# Localization of Excess Temperature Using Plasmonic Hot Spots in Metal Nanostructures: Combining Nano-Optical Antennas with the Fano Effect

Larousse Khosravi Khorashad,[1] Lucas V. Besteiro,[1] Zhiming Wang,[2] Jason Valentine,[3]

Alexander O. Govorov[1]1

[1]Department of Physics and Astronomy, Ohio University, Athens, Ohio 45701, United States

[2]Institute of Fundamental and Frontier Sciences, University of Electronic Science and Technology of China, Chengdu 610054, PR China; State Key Laboratory of Electronic Thin Films and Integrated Devices, University of Electronic Science and Technology of China, Chengdu 610054, PR China

[3]Department of Mechanical Engineering, Vanderbilt University, Nashville, Tennessee 37212, USA

**Abstract**

It is challenging to strongly localize temperature in small volumes because heat transfer is a diffusive process. Here we show how to overcome this limitation using electrodynamic hot spots and interference effects in the regime of continuous-wave (CW) excitation. We introduce a set of figures of merit for the localization of temperature and for the efficiency of the plasmonic photothermal effect. Our calculations show that the temperature localization in a trimer nanoparticle assembly is a complex function of the geometry and sizes. Large nanoparticles in the trimer play the role of the nano-optical antenna whereas the small nanoparticle in the plasmonic hot spot acts as a nano-heater. Under the peculiar conditions, the temperature increase inside a nanoparticle trimer can be localized in a hot spot region at the small heater nanoparticle and, in this way, a thermal hot spot can be realized. However, the overall power efficiency of temperature generation in this trimer is much smaller than that of a single nanoparticle. We can overcome the latter disadvantage by using a trimer with a nanorod. In the trimer assembly composed of a nanorod and two spherical nanoparticles, we observe a strong plasmonic Fano effect that leads to the concentration of optical energy dissipation in the small heater nanorod. Therefore, the power efficiency of generation of local excess temperature in the nanorod-based assembly greatly increases due to the strong plasmonic Fano effect. The Fano heater incorporating a small nanorod in the hot spot has obviously the best performance compared to both single nanocrystals and a nanoparticle trimer. The principles of heat localization described here can be potentially used for thermal photo-catalysis, energy conversion and bio-related applications.

---

[1] Corresponding author: Govorov@ohio.edu; ++1-740-593-9430

**Introduction**

Efficient and localized heat generation using plasmonic nanocrystals and optical excitation is an interesting and active field of research. The motivations for this research direction come from both current and potential applications of optically-excited nanoparticles. These applications include a wide range of bio-related methods and approaches,[1,2,3,4,5] photothermal and hot-electron mechanisms of catalysis,[6,7] thermal actuation of bio-systems,[8] efficient light-to-heat conversion,[9,10] boiling and steam generation,[11,12,13] local temperature sensing in nanoscale systems,[14] and so on.

Optically-excited nanocrystals have large absorption cross sections and, therefore, are able to generate heat very efficiently[15]. Strength and localization of heat generation strongly depend on a geometry and composition of plasmonic nanostructures and, therefore, one can design nanostructures for special photothermal applications. The regimes of photo-heating of a matrix can be collective or local. The collective heating is realized in a large and dense ensemble of nanoparticles (NPs)[16,17,18] where heat fluxes from individual NPs add up leading to high collective temperature. This regime is suitable for the purpose of uniform heating of large areas that can be used in biology, for example. The collective heating is typically realized in the regime of CW-illumination[9,16] when the system has enough time to come to the non-equilibrium steady state with increased temperature. The local regime of heating can provide high local temperatures in certain confined volumes within or in the vicinity of a plasmonic nanostructure.[9,16,27] The local and collective heating mechanisms depend on the composition and dimensions of a complex.[16,17] For some NP assemblies, collective electromagnetic and heating effects can be understood within the effective-medium approaches.[11,19] Moreover, the local temperature distribution and the efficiency of temperature generation depend on the shape of a nanocrystal.[20]

One of the ideas for localized temperature generation is a nanostructure with local electromagnetic hot spots. For the first time, such a nanostructure was proposed in Ref. [16]. The complex proposed in Ref. [16] consisted of three NPs and had a linear geometry. Two large NPs played a role of an antenna and another small NP was placed into the hot-spot region between the two large antenna NPs. As a result, it was calculated[16] that the small NP heater should be able to generate much higher temperature because of the antenna hot-spot effect. However, this problem of the NP-NP-NP heater was not investigated in detail in Ref. [16] and, in general, we will show in our study that this thermal hot-spot problem is not straightforward. Recently, the heterogeneous Ag-Au-Ag trimer was investigated in detail in Refs.[21,22], and it was shown that the temperature generation inside such assembly can be greatly enhanced in the pulsed regime. More studies of photothermal effects in nanocrystal assemblies of various geometries can be found in Refs.[23,24]. In particular, Ref.[24] investigated the photothermal effect in nanorods in the regime of the Fano effect.

Here we study nanocrystal assemblies with hot spots to create strong and localized heating. We compose assemblies from spherical nanoparticles (NPs) and nanorods (NRs). In our geometry, large NPs play a role of an antenna, whereas a small nanocrystal (sphere or nanorod) works as a heater. Due to the electromagnetic hot-spot effect, our structures exhibit strongly-enhanced temperature generation and strongly-localized temperature distributions. The formation of thermal hot spots in our resonant structures requires peculiar conditions that depend on the geometry and sizes of nanocrystals. In particular, we demonstrate the formation of the thermal hot spots in the NP-NP-NP trimers with narrow gaps and for the regime of the Fano resonance in the NP-NR-NP geometry. It is interesting to introduce figures of merit for the photothermal effects in simple and complex nanostructures. Such figures of merit have been already formulated for the efficiency of heat generation using various materials and for the temperature-related optical nonlinearity.[22,25] In our study, we introduce novel types of figures of merit for the spatial localization of temperature and for the power efficiency of localized temperature generation. We then apply our figures of merit to the NP-NP-NP and NP-NR-NP complexes. We show that, under special conditions, the electromagnetic hot-spot effect leads to the formation of thermal hot spots. Here we describe such conditions for the regime of CW-illumination. We should briefly discuss important differences between our present model and the model of the Ag-Au-Ag heater calculated in Ref.[22]. Ref.[22] investigated the pulsed regime and the thermal nonlinearity of the plasmonic response of a small Au-NP. Our paper deals with homogeneous Au-Au-Au complexes in the CW-excitation regime. The pulsed-excitation and CW-excitation regimes have very different properties and, in particular, the degree of localization of temperature can be very different for these regimes.[26,27] Other differences: (1) We describe the thermal effect for the Fano regime with nanorods, when the temperature generation and efficiency become strongly amplified; (2) We introduce the figures of merit for the localization and efficiency of photo-temperature for the plasmonic heaters and test these figures of merit for some important geometries. Finally, the original proposal for the NP-NP-NP plasmonic heater was published in our paper back in 2006[16], but, in the present study, we elaborate on this problem and formulate the conditions for the appearance of thermal hot spots in the CW-illumination regime.

This paper is focused on the CW-illumination regime. The advantages of the CW-regime are obvious: (1) It is easier to realize such regime; (2) This regime is relevant to the energy applications since solar illumination should be considered as continuous. For both excitation regimes (i.e. for the CW- and pulsed illuminations), one can localize temperature using specially-designed nanostructures, but these two regimes may give very different temperature distributions and should be considered separately.

Figure 1 shows the systems studied in this paper. Such nanoscale heater systems can be fabricated using bio-functionalization of NPs or with lithography.[13,28,29] The current literature has a large number of cases related to both approaches. NP and NR trimers can be assembled using the approach of DNA-origami.[30,31,32,33] Figure 1d shows a sketch of a possible DNA-based bio-assembly for photo-heating effect. Plasmonic nanostructured heaters can also be fabricated or arranged on functionalized surfaces using a combination of lithography and bio-assembly.[13,28,34] We finally note that local temperature measurements in a nanoscale heater can be made by attaching a temperature-sensitive emitting nanocrystal[14] to the hot element of assembly.

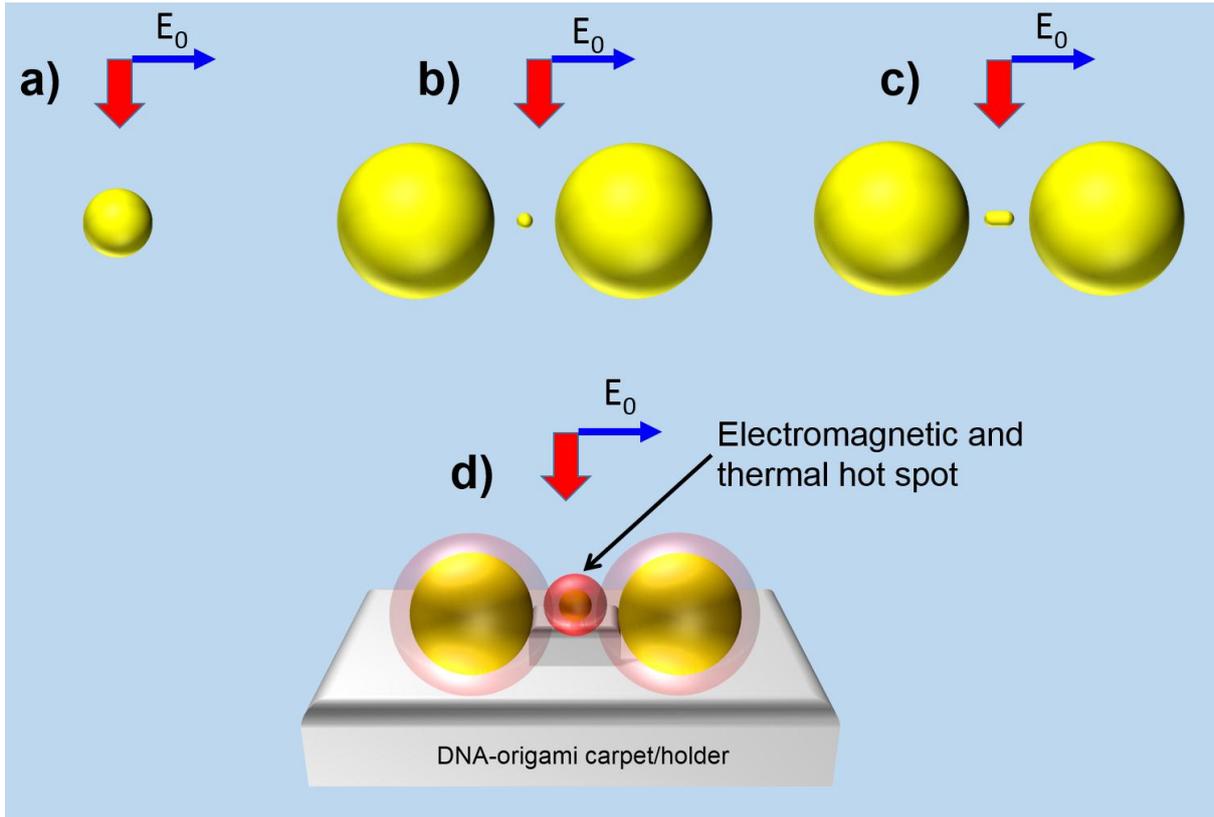

**Figure 1: (a-c)** Thermal and electromagnetic models of nanocrystal trimers with thermal and electromagnetic hot spots. **(d)** Sketch of a structure that can be assembled using the DNA-origami technique. The DNA carpet holds three nanoparticles. The system has an additional small platform that supports a small nanoparticle between two large plasmonic nano-spheres. The system is submerged in water.

1. **General equations for the plasmonic heat generation**

In the following, we will operate with the function $\Delta T(\mathbf{r},t) = T(\mathbf{r},t) - T_0$ that is a local temperature increase with respect to the equilibrium temperature $T_0$. In the absence of phase transformations, heat transfer in a system composed of NPs and a matrix is described by the usual heat transfer equation[16,26]

$$\rho(\mathbf{r}) c(\mathbf{r}) \frac{\partial T(\mathbf{r},t)}{\partial t} = \vec{\nabla} \cdot \left( k(\mathbf{r}) \vec{\nabla} T(\mathbf{r},t) \right) + Q(\mathbf{r},t), \qquad (1)$$

where $T(\mathbf{r},t)$ is the temperature increase as a function of the coordinate $\mathbf{r}$ and the time $t$; $\rho(\mathbf{r}), c(\mathbf{r})$ and $k(\mathbf{r})$ are the mass density, specific heat and thermal conductivity, respectively. The local heat intensity $Q(\mathbf{r},t)$ comes from light dissipation in Au NPs:

$$Q(\mathbf{r},t) = \langle \mathbf{j}(\mathbf{r},t) \cdot \mathbf{E}(\mathbf{r},t) \rangle_t = \frac{\omega}{8\pi} |\mathbf{E}_\omega(\mathbf{r})|^2 \cdot \operatorname{Im} \varepsilon(\mathbf{r}), \qquad (2)$$

where $\mathbf{j}(\mathbf{r},t)$ is the current density, $\mathbf{E}(\mathbf{r},t) = \operatorname{Re}\left[ \mathbf{E}_\omega(\mathbf{r}) \cdot e^{-i\omega t} \right]$ is the resulting electric field in the system and $\varepsilon(\mathbf{r})$ is the local dielectric constant of the system. Here we assume that the system is excited with the monochromatic external laser field, $\mathbf{E}_0(\mathbf{r},t) = \mathbf{E}_0(t) \cdot \operatorname{Re}\left[ e^{-i\omega t + i\mathbf{k}\cdot\mathbf{r}} \right]$, where $\mathbf{E}_0(t)$ is the amplitude of the electric field in the electromagnetic wave. Then, the light intensity as a function of time is given by $I(t) = c_0 \mathbf{E}_0(t)^2 \sqrt{\varepsilon_0} / 8\pi$, where $c_0$ is the speed of light in vacuum. For example, if light turns on at $t=0$, the light intensity becomes: $I_{flux} = c_0 \mathbf{E}_0^2 \sqrt{\varepsilon_0} / 8\pi$ for $t>0$ and $0$ at $t<0$. Here $\varepsilon_0$ is the dielectric constant of the matrix. For the thermal coefficients of water and gold, we will use the standard literature numbers given in Table 1. For the electrodynamic parameters, we use the data from Palik for gold[35] and the matrix constant of $\varepsilon_0 = 1.8$. This matrix dielectric constant corresponds to water. If a frame of a structure is made of DNAs or polymers, effective dielectric constants of such molecules are also close to the one of water.

$$
\begin{aligned}
&\textit{Water}: \\
&K_{diff,w} = \frac{k_{t,w}}{\rho_w \cdot c_w} = 0.143 \cdot 10^{-6} \frac{m^2}{s} \\
&k_{t,w} = 0.6 \frac{W}{m \cdot K} \\
&\rho_w = 10^3 \frac{kg}{m^3} = 1 g/cm^3 \\
&c_w = 4181 \frac{J}{kg \cdot K} \\
&\textit{Gold}: \\
&K_{diff,Au} = \frac{k_{t,Au}}{\rho_{Au} \cdot c_{Au}} = 128 \cdot 10^{-6} \frac{m^2}{s} \\
&k_{t,Au} = 318 \frac{W}{m \cdot K} \\
&\rho_{Au} = 19300 \frac{kg}{m^3} = 19.3 \frac{g}{cm^3} \\
&c_{Au} = 129 \frac{J}{kg \cdot K}
\end{aligned}
$$

Table 1: Thermal coefficients of water and gold used in the calculations. The thermal diffusivity of a material is related to the thermal conductivity via $K_{diff} = k_t / (\rho \cdot c)$.

2. **Localization of temperature, spatial variations of temperature and the related figures of merit for thermal hot spots**

How to characterize the degree of localization of phototemperature near optically-excited plasmonic NPs and, in general, spatial variations of temperature in complex systems? For this purpose, we propose the following simple figures of merit. Figure 2 illustrates these physical parameters. The first figure of merit is the localization length of temperature,

$$\alpha_{localization\ length} = \frac{\Delta L_{heating}}{L_{heater}}. \qquad (3)$$

Here $L_{heater}$ is the size of a plasmonic heater, which can be a single NP or a NP complex. The length $\Delta L_{heating}$ is the dimension of the heated area where the local temperature increase $\Delta T(\mathbf{r})$ is above the value of $\Delta T_{max}/2$. We note that the maximum temperature, $T_{max}$, can occur at a surface of one of metal NPs in our system. This is because of a very high heat conductivity of a metal. In our cases, we will always deal with gold (Table 1). The second figure of merit should describe the ability of a plasmonic structure to create large temperature gradients,

$$\alpha_{temperature\ gradient} = \frac{\left|\frac{dT(\mathbf{r})}{dl}\right|_{max}}{I_{flux}}, \qquad (4)$$

where $\left|dT(\mathbf{r})/dl\right|_{max}$ is the maximum temperature gradient anywhere in the system and $I_{flux}$ is the light flux. We note that the parameter (3) is dimensionless and both parameters (3) and (4) are independent of the incident light power. These parameters depend on architecture of a heater and on thermal and electromagnetic parameters of a hybrid system. We now introduce another type of figures of merit, the relative efficiencies of generation of local excess temperature. These efficiencies can be defined as

$$Eff_{temperature-flux} = \frac{\Delta T_{max}}{I_{flux}},$$

$$Eff_{temperature-absorption} = \frac{\Delta T_{max}}{Q_{absorption,tot}}, \qquad (5)$$

where $Q_{absorption, tot}$ is the rate of total absorption in the plasmonic system in the CW-excitation regime. We will see that, for most interesting regimes of heating in our trimer structures, the maximum temperature $\Delta T_{max}$ appears at the surface of a small NP. In such cases, the parameters (5) should include $\Delta T_{max} = \Delta T_{at\,small\,NP}$. We note that the parameters (5) can also be introduced for the pulsed regime. For the pulsed regime, we should define $\Delta T_{max}$ as the maximum temperature increase achieved during the pulse interval $\Delta t_{pulse}$. In addition, $I_{flux}$ and $Q_{absorption, tot}$ should be regarded as the flux and the heat rate during the pulse. Then, the total illumination power density should be defined as $P_{tot} = I_{flux} \cdot \Delta t_{pulse}$ and the total absorbed energy should be calculated as $Q_{absorption, tot} \cdot \Delta t_{pulse}$.

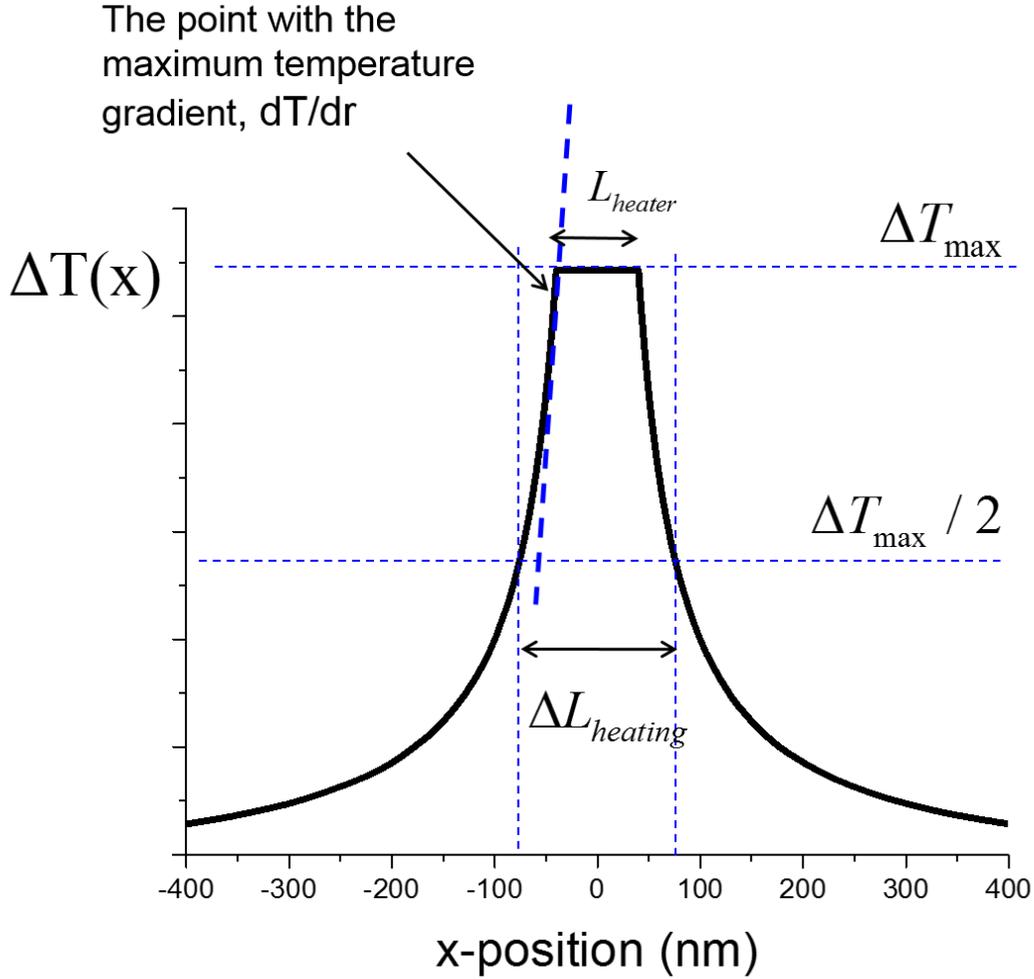

**Figure 2:** Generic non-uniform temperature distribution around a nanoscale heater and the parameters involved in the spatial figures of merit.

3. **Solutions and the figures of merit for a single nanoparticle heater.**

We now analyze the above figures of merit for the simplest case of a single optically-excited NP. This case of a single NP will serve as a reference for more complex NP assemblies. For a single NP, we can write analytical equations for the CW- and pulsed-illumination regimes. The CW-illumination regime has the following temperature distribution[16,26]:

$$\Delta T(\mathbf{r}) = \begin{cases} \dfrac{Q_{tot}}{4\pi k_{t,w}} \dfrac{1}{r} = \Delta T_{max} \dfrac{R_{NP}}{r} & r > R_{NP} \\ \dfrac{Q_{tot}}{4\pi k_{t,w}} \dfrac{1}{R_{NP}} = \Delta T_{max} & R_{NP} > r > 0 \end{cases}, \quad (6)$$

where $R_{NP}$ is the NP radius and $k_{t,w}$ is the thermal conductivity of water. For the metal, $k_{t,Au} \gg k_{t,w}$ and temperature inside a NP can be taken constant. It follows from Eq.6 that the maximum temperature increase in the light-driven system, which is also the temperature at the NP surface, is given by the simple equation

$$\Delta T_{max} = \frac{Q_{tot}}{4\pi k_{t,w}} \frac{1}{R_{NP}}. \quad (7)$$

For small NPs, the total rate of heat dissipation $Q_{tot}$ depends on the induced electric field inside an Au-NP and is given by the analytic equation

$$Q_{tot} = V_{NP} \frac{\omega}{8\pi} \mathbf{E}_0^2 \left| \frac{3\varepsilon_0}{2\varepsilon_0 + \varepsilon_{Au}} \right|^2 \operatorname{Im}\varepsilon_{Au}(\mathbf{r}), \quad (8)$$

where $\varepsilon_{Au}(\omega)$ and $\varepsilon_0$ are the dielectric constants of gold and water, respectively; $V_{NP}$ is the NP volume. The dissipation given by Eq. 8 depends on the electric field amplitude that is a function of the light intensity, $\mathbf{E}_0^2 = \dfrac{8\pi}{c_0\sqrt{\varepsilon_0}} I_0$. For large NPs, the near-field equation (8) is no longer valid, and we need to use the standard Mie theory.[36] Figure 3 shows the numerical results for a single NP in water. We used COMSOL to obtain the electromagnetic and thermal properties. In particular, we see that $\Delta T_{max}(a_{NP}) \propto a_{NP}^2$ for small NP sizes ($a_{NP} \ll \lambda$), as shown in Figure 3d. Here $a_{NP}$ denotes the NP diameter and, correspondingly, $R_{NP} = a_{NP}/2$. This limit is consistent with Eqs. 7 and 8. For larger NPs, the electromagnetic retardation effects become strong and the dependence $\Delta T_{max}(a_{NP})$ becomes more complex. In particular, the absorption cross section is not proportional to the NP volume anymore (Figure 3d).

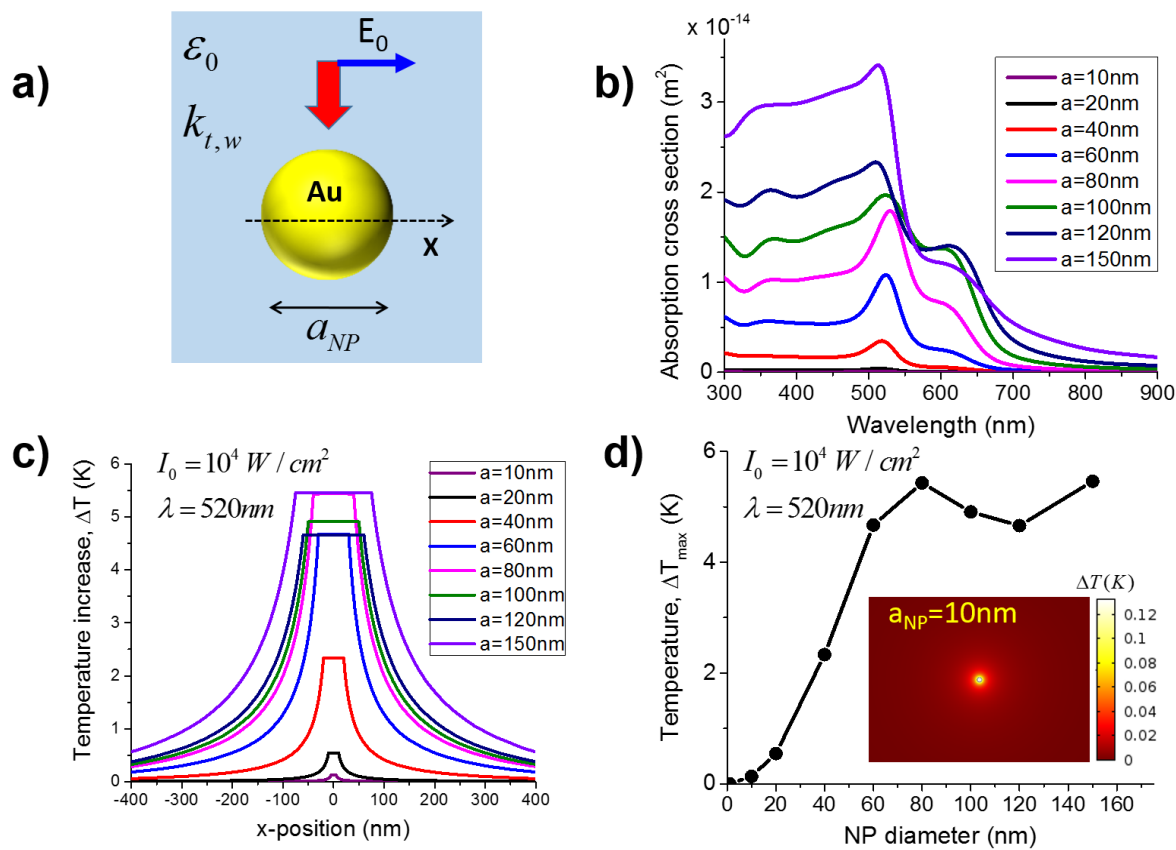

**Figure 3:** Computer simulations of the electromagnetic and thermal properties of a single NP heater. **(a)** Model of a single Au-NP in a thermally-conductive matrix. **(b,c)** Absorption cross sections and spatial temperature distributions for single Au-NPs. **(d)** Temperature at the surface of a single NP as a function of the NP diameter. **Inset:** Temperature map for a 10nm-NP in water.

Then, for a single NP heater, the localization length is given apparently by $2a_{NP}$ and the maximum temperature gradient occurs near the NP surface in the matrix at $r \to R_{NP}$. This gradient is given by $\left|\frac{dT(\mathbf{r})}{dx}\right|_{max} = \frac{\Delta T_{max}}{R_{NP}}$. Then, the figures of merit for the CW-regime become:

$$\alpha_{localization\ length} = \frac{\Delta L_{heating}}{L_{heater}} = 2,$$

$$\alpha_{temperature\ gradient} = \frac{\left|\frac{dT(\mathbf{r})}{dl}\right|_{max}}{I_{flux}} = \frac{\Delta T_{max}}{R_{NP} I_{flux}}.$$

Note that the second parameter $\alpha_{temperature\ gradient}$ has the size dependence. For small NPs, $\alpha_{temperature\ gradient}^{static} \propto R_{NP}$. For larger NPs, we can create stronger temperature variations since the temperature generation grows with the size as $\Delta T_{max} \propto R_{NP}^2$. Since the above figures of merit belong to the simplest heating system, these figures can be used as references for more complex nanoscale heaters. Now we look at the power parameters of merit for a single NP:

$$Eff_{temperature-flux} = \frac{\Delta T_{max}}{I_{flux}} = \frac{1}{4\pi k_{t,w}} \frac{c_{abs}}{R_{NP}},$$

$$Eff_{temperature-absorption} = \frac{\Delta T_{max}}{Q_{absorption,tot}} = \frac{1}{4\pi k_{t,w}} \frac{1}{R_{NP}},$$

where $c_{abs} = Q_{tot}/I_{flux}$ is the absorption cross section of a NP. These parameters have units and should be calculated numerically for a single NP and for NP complexes. For a single NP of small size, these parameters have simple analytical expressions and strongly depend on the size:

$$Eff_{temperature-flux} \propto R_{NP}^2,$$

$$Eff_{temperature-absorption} \propto \frac{1}{R_{NP}}.$$

In the pulsed regime, the general solution for a small NP is more complex and, therefore, we consider below two temporal regimes.

**Pulsed excitation, short times**. For short times, the temperature distribution $\Delta T(\mathbf{r},t)$ is given by a special function.[26] However, we can make the following approximations for the localization length: (1) The heat diffusion distance from the NP surface can be estimated by the diffusion length, $\delta L(t) = \sqrt{K_{diff,w} t}$, where $t$ is time; (2) Then, the size of the hot spot can be approximated as $\Delta L_{heating}(t) \sim L_{heater} + 2 \cdot \delta L(t)$. Therefore, the dynamic figures of merit get the following approximations:

$$\alpha_{localization\ length}^{dynamic} = \frac{\Delta L_{heating}}{L_{heater}} \sim 1 + \frac{\sqrt{K_{diff,w} t}}{R_{NP}}$$

$$\alpha_{temperature\ gradient}^{dynamic} = \frac{\left|\frac{dT(r)}{dl}\right|_{max}}{I_{flux}} \sim \frac{\Delta T_{max}(t)}{I_{flux}} \frac{1}{\sqrt{K_{diff,w} t}}$$

We note that the above dynamic parameter did not become infinite at very small times since all these equations are valid for $t > \Delta t_{pulse}$. We see that the pulsed regime has obviously an advantage for short times when the localization is strong (small parameters $\alpha_{localization\ length}^{dynamic}$) and the temperature gradients are at maximum (large parameters $\alpha_{temperature\ gradient}^{dynamic}$). In other words, the dynamic localization parameter is smaller than the corresponding static one for short times: $\alpha_{localization\ length}^{dynamic} < \alpha_{localization\ length}^{static} = 2$. This is because of the finite time for heat to diffuse from a NP. The same argument is applied for the thermal-gradient parameter, $\alpha_{temperature\ gradient}^{dynamic} > \alpha_{temperature\ gradient}^{static}$. The pulsed regime is able to produce stronger temperature gradients.

**Pulsed excitation, long times.** For long times, when the heated spot becomes large, we can view the NP as a delta-functional heater. The corresponding condition for the time is $t \gg R_{NP}^2 / K_{diff,w}$. For NPs with diameters of 10nm and 100nm, the times should be $t \gg 2\,\text{ns}$ and $t \gg 200\,\text{ns}$, respectively. Now we look at the short-pulse regime assuming $Q(\mathbf{r},t) = P_{tot} \cdot \delta(t) \cdot \delta(\mathbf{r})$, where the total absorbed energy is $e_{tot} = I_{flux} \Delta t_{pulse}$ and $\Delta t_{pulse} \to 0$. The solution for Eq. 1 is well-known[26]:

$$\Delta T(\mathbf{r},t) = \frac{e_{tot}}{\rho_w c_w} \cdot \left( \frac{1}{2\sqrt{\pi K_{diff,w} t}} \right)^3 e^{-\frac{r^2}{4K_{diff,w}t}}.$$

Then localization length obviously depends on time, $\Delta L_{heating} = 2\sqrt{Ln(2) \cdot K_{diff,w} \cdot t}$. And the time-dependent figures of merit are

$$\alpha_{localization\ length}^{dynamic} = \frac{\Delta L_{heating}}{L_{heater}} = \frac{\sqrt{Ln(2) \cdot K_{diff,w} \cdot t}}{R_{NP}},$$

$$\alpha_{temperature\ gradient}^{dynamic} = \frac{\left|\frac{dT(\mathbf{r})}{dx}\right|_{max}}{I_{flux}} = \frac{const}{\left(K_{diff,w} \cdot t\right)^2}.$$

We see that, for long times, the figures of merit for the localization, $\alpha_{localization\ length}^{dynamic}$, increases that manifests an increase of the localization region. Simultaneously, the thermal-gradient parameter, $\alpha_{temperature\ gradient}^{dynamic}$, decreases for the same reason of spreading of the heated spot due to diffusion. In other words, as expected, the temperature distribution at long times becomes less localized and, simultaneously, the temperature gradients decrease with time and eventually disappear for $t \to \infty$.

To conclude, compared to the CW-regime, the pulse excitation creates overall a more strongly-localized temperature distributions at short times since the temperature increase drops exponentially as a function of the radial coordinate. However, the degree of localization at long times becomes strongly reduced because of spreading of the heated spot.

**4. The NP-NP-NP assembly with a hot spot for localized heating**

Figures 1b and 4 show the geometry with three spherical NPs. The simple idea behind this structure was already pointed out in our old paper,[16] but it was not investigated at that time in details. The idea is to use two large NPs as a nano-optical antenna to enhance the electromagnetic field and heating at the small heater NP. The sizes of larger antenna NPs are chosen as 100nm, whereas the small spherical heater has a smaller dimeter of 10nm. The NP-NP antenna with the chosen sizes is able to strongly enhance the electromagnetic field in the center (Figure 5). Regarding the temperature generation, the challenge here is the expected strong dissipation rate in the large 100nm-NPs. This dissipation can create a strong temperature increase at the surfaces of large NPs and it also can limit the power efficiencies of temperature generation at the small 10nm-NP. If such, the hot-spot generation of temperature in the small NP would become a secondary effect. Indeed our calculations show different regimes of temperature generation that depend strongly on the sizes of large NPs and on the distances in a trimer. Below we discuss these regimes in detail.

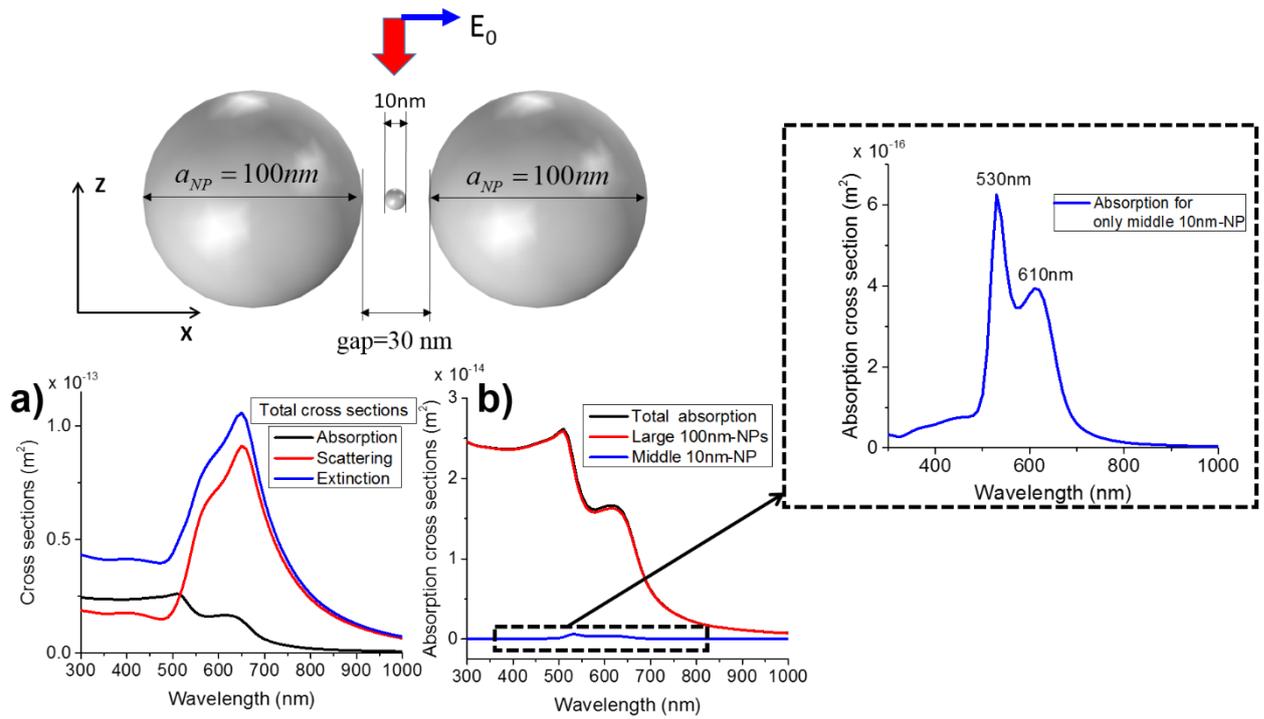

**Figure 4:** Computer simulations of the electromagnetic properties of the NP-NP-NP trimer. **Insets**: Model of the trimer system and the plot for the resonant absorption region of the 10nm-NP.

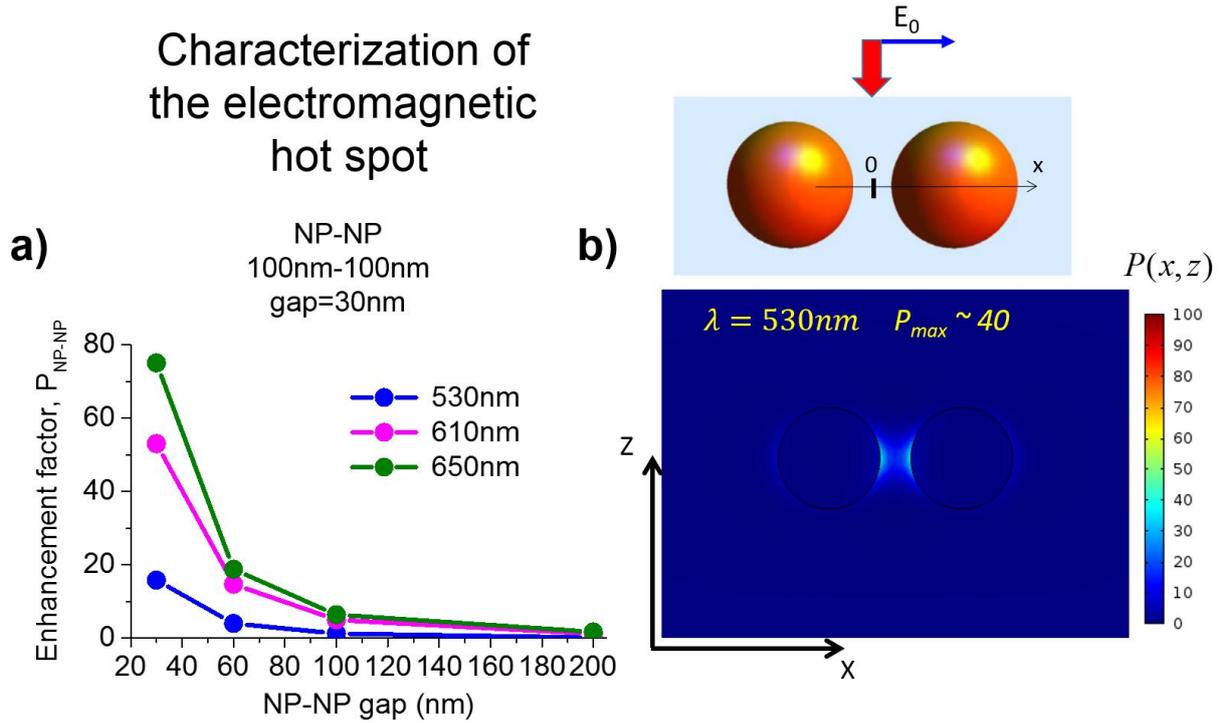

**Figure 5:** Electromagnetic properties of the NP-NP antennas without the central NP. **(a)** Field enhancement factor in the center of the dimer for the three wavelengths used in the temperature calculations. **(b)** Spatial map of the field enhancement factor with the electromagnetic hot spot in the center of the NP-NP antenna with *gap*=30nm. **Inset:** Model and geometry.

First of all, we look at the heat powers generated by the whole complex and by the small 10nm-NP inside the complex (Figure 4). We see that the heat generation by the small 10nm-NP heater has the resonances at 530nm and 610nm. At these resonances, we now compute the temperature distributions inside the complex (Figures 6 and 7). And we observe rather interesting patterns of the temperature distribution that can be understood qualitatively. In particular, the system can exhibit a temperature spike at the small NP for some geometrical parameters. The spike temperature, $\delta T_{peak}$, is defined in Figure 6c. According to Figure 3d, an isolated small 10nm-NP generates a lower temperature increase compared to any other single large NP in our complexes. The simple dependence for the generation of temperature for small NPs is: $\Delta T_{max} \propto R_{NP}^2$.[16,17] Then, the small 10nm-NP in the trimer generates the following temperature spike in the middle of the system

$$\delta T_{peak} \propto R_{NP}^2 \cdot P_{NP-NP} \cdot I_{flux}, \quad (9)$$

where $P_{NP-NP} = P(r=0)$ is the field enhancement factor in the center of the antenna NP-NP complex. The position-dependent field enhancement factor $P(\vec{r})$ is defined as

$$P(\vec{r}) = \frac{\mathbf{E}_\omega \cdot \mathbf{E}_\omega^*}{E_0^2}.$$

The field enhancement factor in the centre of the antenna NP-NP complex is plotted in Figure 5 for the three resonance frequencies. Of course, the factor $P_{NP-NP}$ decreases with the NP-NP gap in the antenna NP-NP dimer (Figure 5). We note that the above factor $P_{NP-NP}$ was computed for the dimer NP-NP structure to reveal the strength of enhancement. Therefore, the enhancement effect for the temperature spike in the middle of the nano-heater system should diminish and then should vanish for large gaps, $d$. Indeed, we observe this behaviour for the gaps $d > 50\text{nm}$ (Figure 7a). For the most interesting case of small gaps ($d \sim 30\text{nm}$), the temperature distribution (Figure 7a) is special, exhibiting the temperature spike in the middle. This behaviour can be understood in the following way: The two large 100nm-NPs create a large temperature increase of about 5K at their surfaces since we deal with large plasmonic NPs. Then, the small 10nm-heater in the middle can noticeably contribute to the temperature distribution only if the NP-NP antenna creates strong plasmonic field enhancement (see Eq. 9). Without the antenna effect, a single 10nm-NP cannot create a large temperature increase for the given light intensity simply because this heater NP is small in size. In addition, the temperature is almost constant inside the large 100nm-NPs because of the very high thermal conductivity of gold. Then, the resulting temperature profile for the small gaps ($d \sim 30$nm) has a spike in the centre and two flat regions on the sides. For the gaps $d > 50$nm, the central spike almost vanishes since the small 10nm-NP cannot generate much heat without a strong antenna effect (Figure 7a). In Figure 7, we summarize the data for the NP-NP-NP trimer with the antenna effect. Overall, we found that the temperature distributions depend critically on the sizes of the larger NPs (antenna NPs) and on the size of the central NP-heater. The temperature focusing effect is non-trivial since the antenna itself, which is made of large NPs, is an efficient generator of photo-heat.

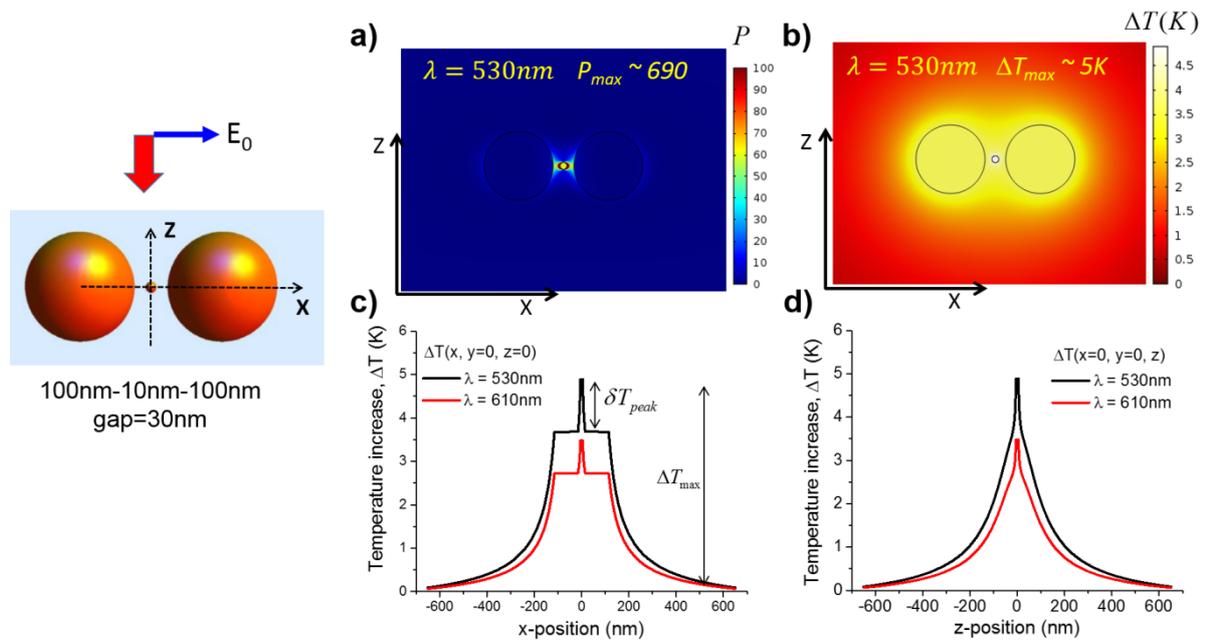

**Figure 6:** Computer simulations of the field and temperature distributions in the optically-excited trimer with gap=30nm. **(a,b)** Electromagnetic and temperature maps. **(c,d)** the temperature distributions along the two axes. **Inset**: Model.

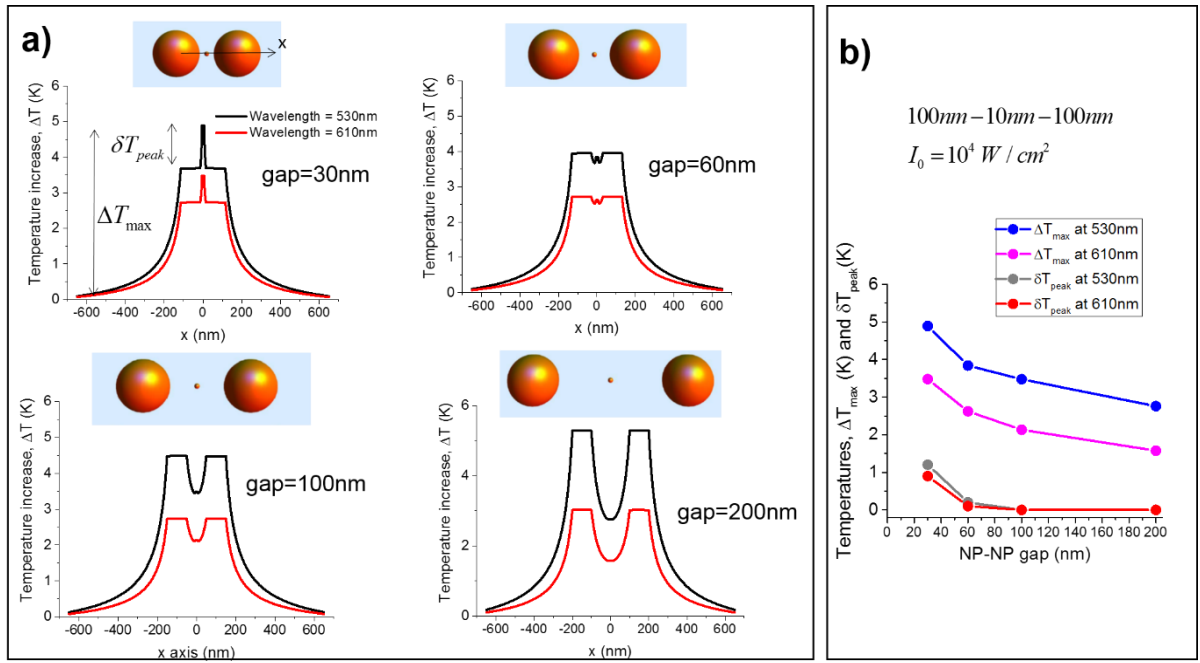

**Figure 7:** **(a)** Computer simulations of the temperature distributions in optically-excited trimers with different gaps, d=30, 60, 100 and 200nm. **(b)** Characteristic temperatures in the thermal hot-spot region of the trimer. **Insets**: Models.

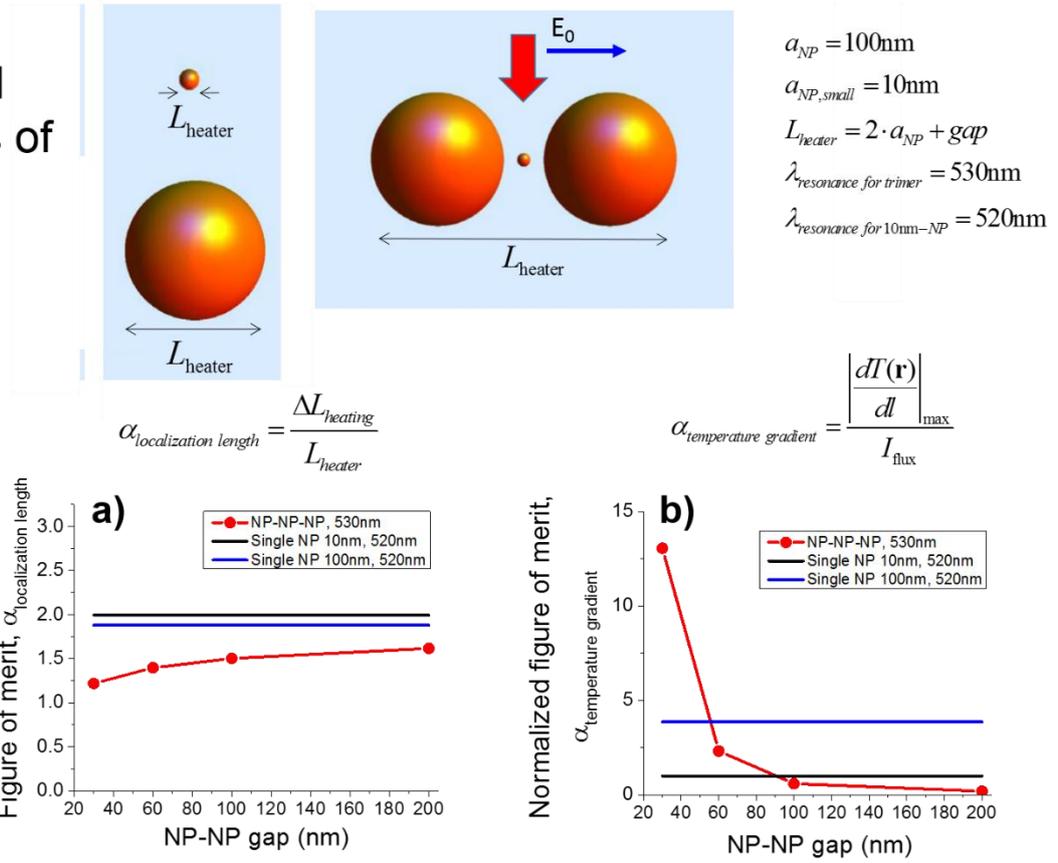

**Figure 8: (a, b)** Spatial figures of merit for the NP-NP-NP trimer. **Insets**: Models.

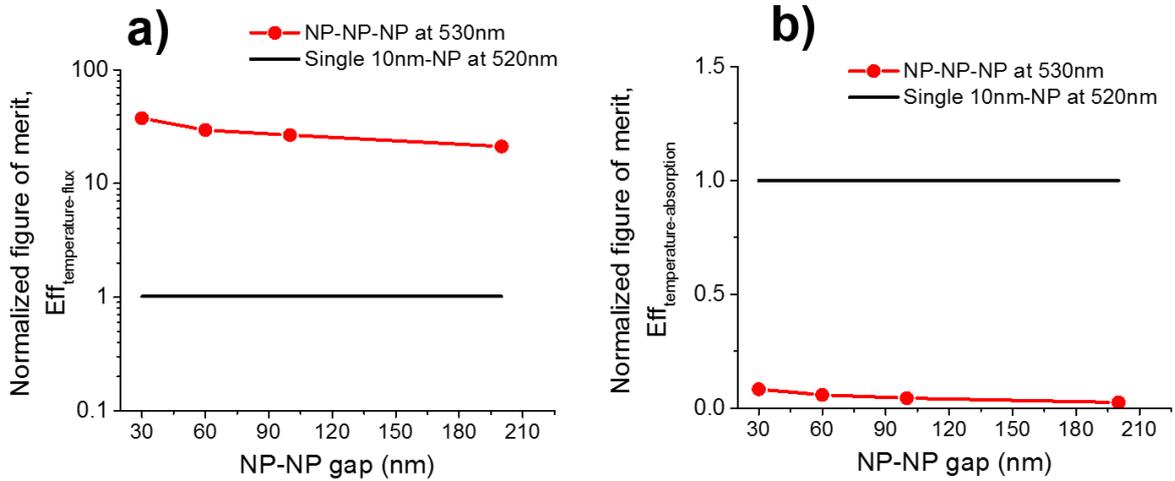

Figure 9: (a, b) Power figures of merit for the NP-NP-NP trimer. **Inset**: Model.

Figures 8 shows the spatial figures of merit at the resonance wavelengths of single NPs and NP-NP-NP trimers. We look now at the figure of merit for the localization length. The general principle is that heater systems with smaller $\alpha_{localization\ length}$ have advantage since the temperature distribution in such systems is more localized. For small and large spherical NPs, $\alpha_{localization\ length} \approx 2$. For the NP-NP-NP trimers, we see that the localization figure of merit, $\alpha_{localization\ length}$, is slightly smaller than that for single NPs (Figure 8a) and, therefore, the trimer is slightly better for localization of temperature. Overall, we see that the trimer structure does not give a real advantage for localization of temperature as compared to a single NP. Then, we look at the temperature-gradient parameter (Figure 8b). We see in Figure 8b that the parameter of temperature gradient ($\alpha_{temperature\ gradient}$) for the small trimers significantly exceeds the corresponding parameters for the small and large single NPs. This is due to the strong temperature gradients in the central region of the trimer between the 10nm-NP and the 100nm-NPs. We see that the tightly-packed trimers have advantage as compared to single NPs. The physical reason is in the electromagnetic focusing of optical energy into the hot spot region of the NP-NP antenna where we place the small 10nm-heater.

In Figure 9, we show the situation with the power figures of merit. The NP-NP-NP trimer performs much better than the single NP heaters for the power parameter $Eff_{temperature-flux}$. This is due to the focusing of optical energy on the small NP. However, the NP-NP-NP trimer cannot compete with a single small-NP heater for the parameter $Eff_{temperature-absorption}$. The reason is, as we mentioned above, strong absorption by the NP-NP antenna. In other words, the NP-NP-NP trimer gives great advantage for a given light flux, but, simultaneously, the trimer consumes lots of energy and it cannot compete with a single small NP. In the next section, we will show how we can overcome this disadvantage of strong dissipation in the large antenna NPs using the Fano interference effects.

## 5. The Fano heater assembled from two spherical NPs and a nanorod

The Fano effect in optics and plasmonics appears when an electronic excitation with a narrow absorption line interacts with another broad resonance.[24,37,38] In our model, we now include a nanorod (NR) with a strong and narrow plasmon resonance and a NP dimer with broad resonances (Figure 10). The resulting NP-NR-NP trimer exhibits the Fano effect seen as a dip in the extinction spectra (Figure 10). The striking feature of the Fano effect is that the extinction of the complex (NP-NR-NP) is not a simple sum of the extinctions of the isolated components (NR and NP-NP antenna). We illustrate this feature in Supporting Information. The appearance of the Fano effect in the extinction spectrum of our complex is expected since, in our system, the narrow plasmon line of the NR should strongly interact with the broad plasmonic band of the NP-NP-antenna. Then, we look at the absorptions of the components (Figure 10b). Importantly, the dip in the absorption cross section of the antenna NPs coincides with the maximum of the absorption by the central NR (Figure 10b). The central NR plays the role of a heater and it receives much optical energy in our structure. This behavior appears at the wavelength of 650nm. Another, obvious advantage of the NR heater is the narrow plasmonic peak.

We now consider the ability of this cluster to generate high photo-temperature (Figures 11, 12 and 13). First, we observe a very strong effect of electromagnetic enhancement for the local temperature at the NR for the trimer system (Figure 11). This effect comes from the electromagnetic hot spot in the NP-NP antenna and from the narrow plasmon resonance of the small NR. Simultaneously, we observe the interference Fano effect that slightly suppresses light dissipation in the NP-NP antenna dimer at 650nm (Figure 10b). These absorption properties affect the magnitude of the local temperature increase near the center of the NP-NR-NP structure (Figures 11 and 12).

The effect of the temperature generation at the NR in the NP-NR-NP complex is remarkably strong. We see that the enhancement factor for temperature is $P_T = \Delta T_{\max,NP-NR-NP} / \Delta T_{\max,NR} \sim 20$. This is due to the electromagnetic hot spot effect in the NP-NP antenna. We then determine the field enhancement factor of the hot spot for our conditions (650nm and gap=40nm) that is $P_{NP-NP} \sim 55$ (Figure 5). This number is larger than the calculated temperature enhancement $P_T \sim 20$. This is expected since the NR is located closely to the large 100nm-NPs, which have lower temperatures, and, therefore, the system should have strong heat currents from the hot NR to the colder 100nm-NPs. This should lower to some degree the temperature of the hot NR.

We now discuss the figures of merit for the Fano heater. Figures 12b and c show the data. First, the degree of localization of temperature is greatly enhanced in the NP-NR-NP complex (Fig. 12b), i.e. the localization parameter $\alpha_{localization\,length}$ for this system gets very small. This tells us that the degree of localization of temperature is very strong. Second, the temperature-gradient parameter ($\alpha_{temperature\,gradient}$) for the Fano NP-NR-NP complex greatly exceeds the corresponding parameters for the NR and NP-NP systems. We see that the Fano heater has clear advantage. Now we look at the power figures of merit shown in Figure 13. The power figure of merit responsible for the amplitude of photo-temperature, $Eff_{temperature-flux}$, becomes strongly enhanced in the case of NP-NR-NP system (Figure 13a). We see that this figure of merit of the NP-NR-NP system (Figure 13a) is much larger than those for a single NR, NP-NP dimer and NP-NP-NP trimer. Also, we see that the NP-NR-NP system has a relatively large absorption figure of merit, $Eff_{temperature-absorption}$ (Figure 13b) despite the large absorption rate of the antenna 100nm-NPs. Then, the parameters $Eff_{temperature-absorption}$ for the single NR system and for the NP-NR-NP complexes are comparable (Figure 13b). We note that, in the previous case of the NP-NP-NP system, the absorption parameter for the NP-NP-NP is much smaller than the corresponding parameter for the single small NP (Figure 13c). In the NP-NR-NP system, the absorption parameter, $Eff_{temperature-absorption}$, is strongly increased due to the narrow plasmonic resonance of NR. In other words, the narrow plasmon resonance of the NR leads to strong localized absorption by the NR at 650nm and, simultaneously, the NP-NP antenna does not exhibit strong absorption at this wavelength. In addition, we see that overall the Fano complex has larger field-enhancement factors (Figure 11) as compared to the case of the NP trimer (Figure 6a). Interestingly, this regime of localized photo-heating in the NP-NR-NP complex appears simultaneously with the Fano effect when the narrow NR resonance strongly interacts with the broad plasmonic band of the 100nm-NPs. To illustrate these interesting behaviors in terms of the figures of merit, we summarize the results for the NP-NR-NP and NP-NP-NP complexes in Figure 13.

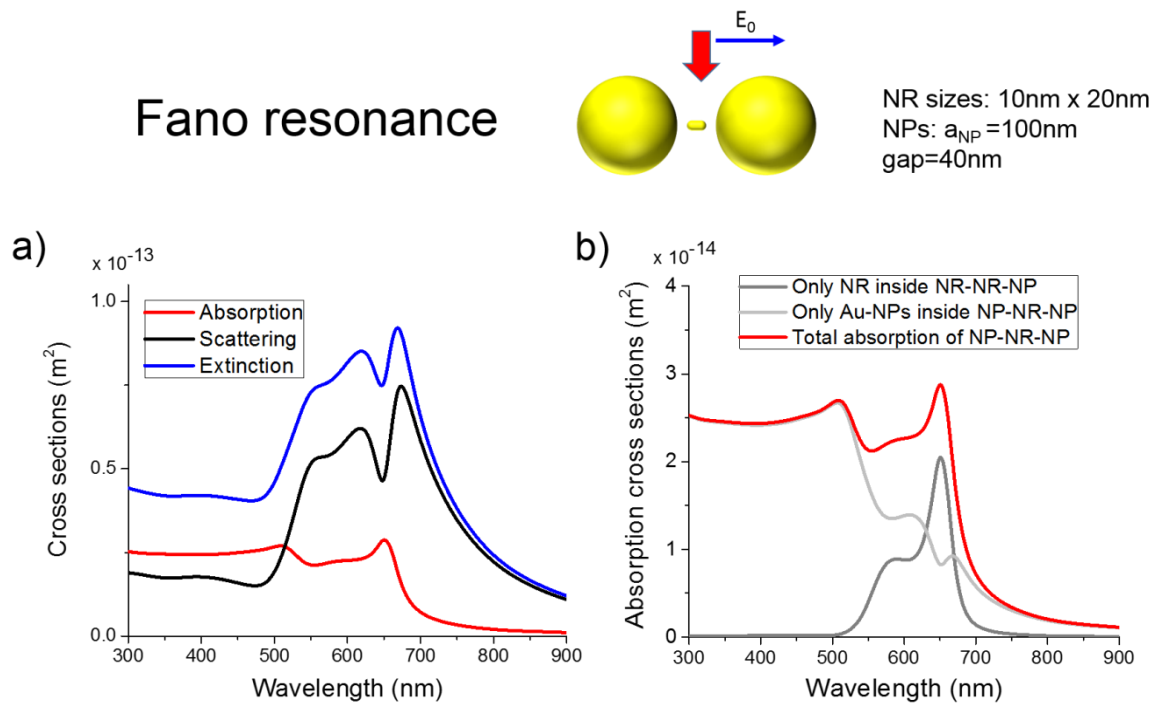

**Figure 10:** Computer simulations of the electromagnetic properties of the NP-NR-NP trimer. **(a)** Optical cross sections of the NP-NR-NP complex. **(b)** Absorption cross sections for the elements within the NP-NR-NP complex. The sum of the absorption cross sections of the NR and the NP-NP antenna gives the total absorption. **Inset**: Model.

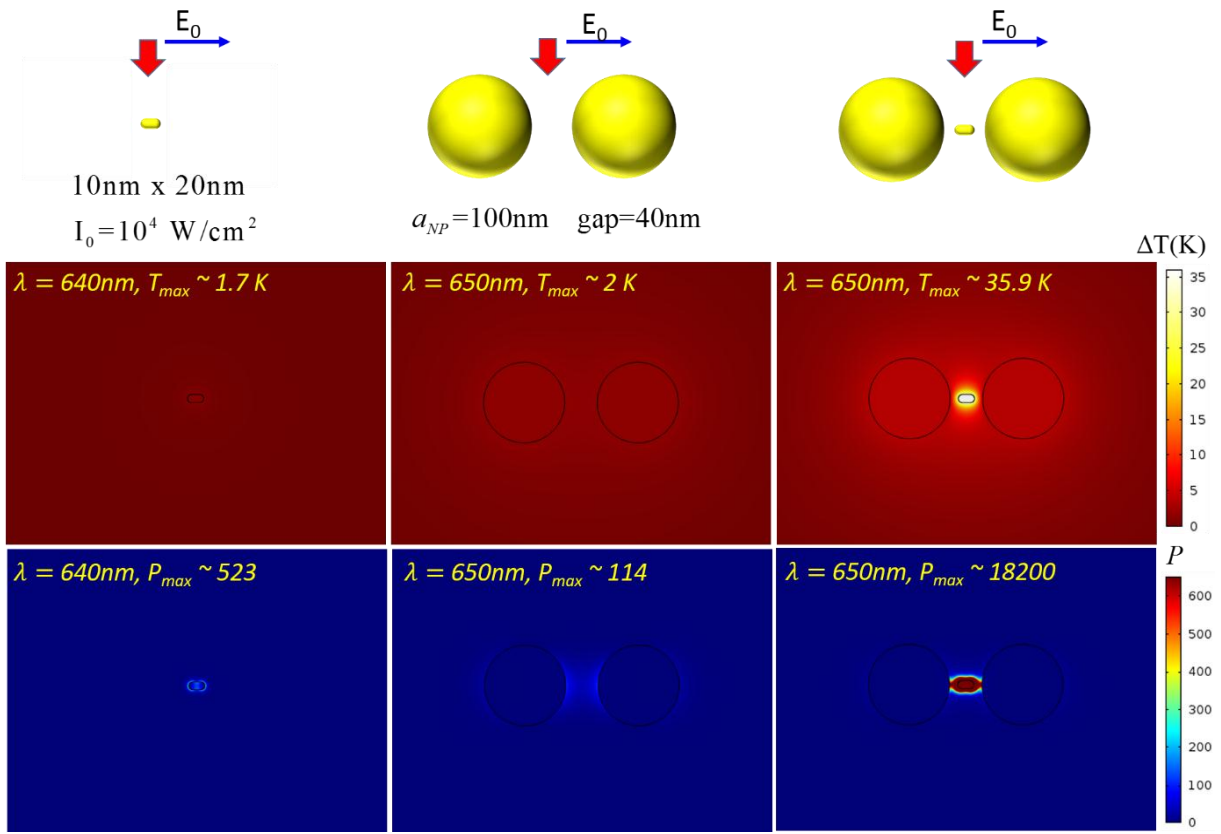

**Figure 11:** Spatial z-x maps of the photothermal properties for the three systems: NR, NP-NP and NP-NR-NP. **Insets**: Models.

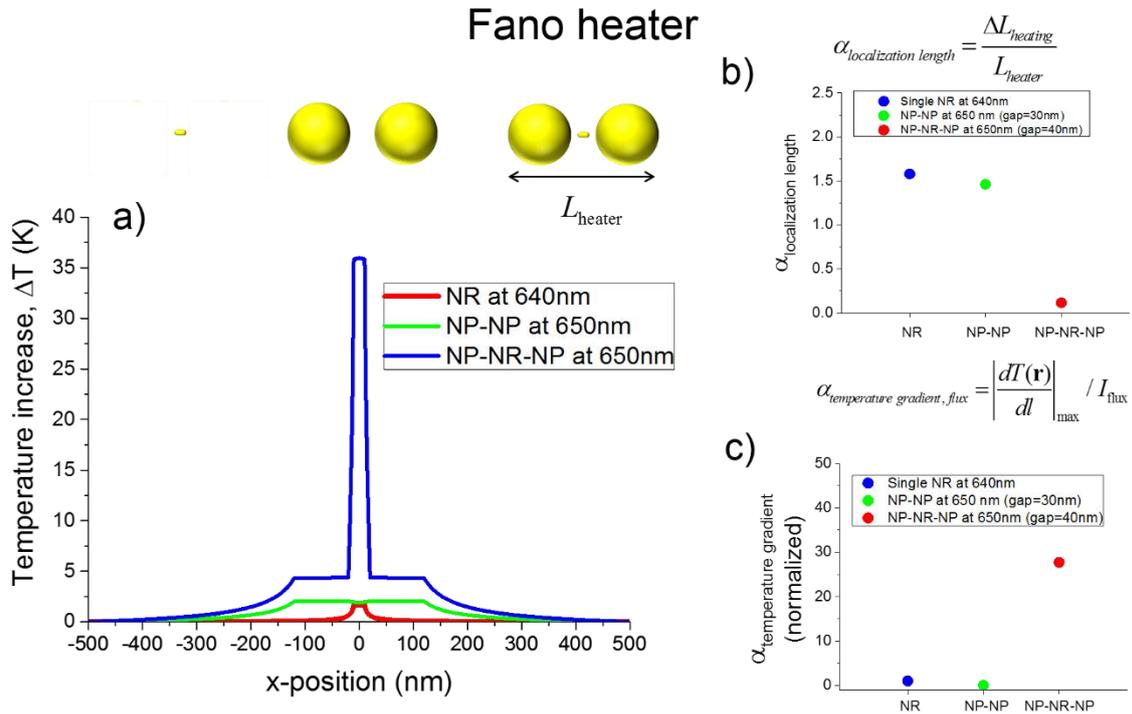

**Figure 12:** **(a)** Computer simulations of the local temperatures for the three systems under study: NR, NP-NP and NP-NR-NP. **(b, c)** Spatial figures of merit calculated for the Fano heater and related structures. **Insets**: Models.

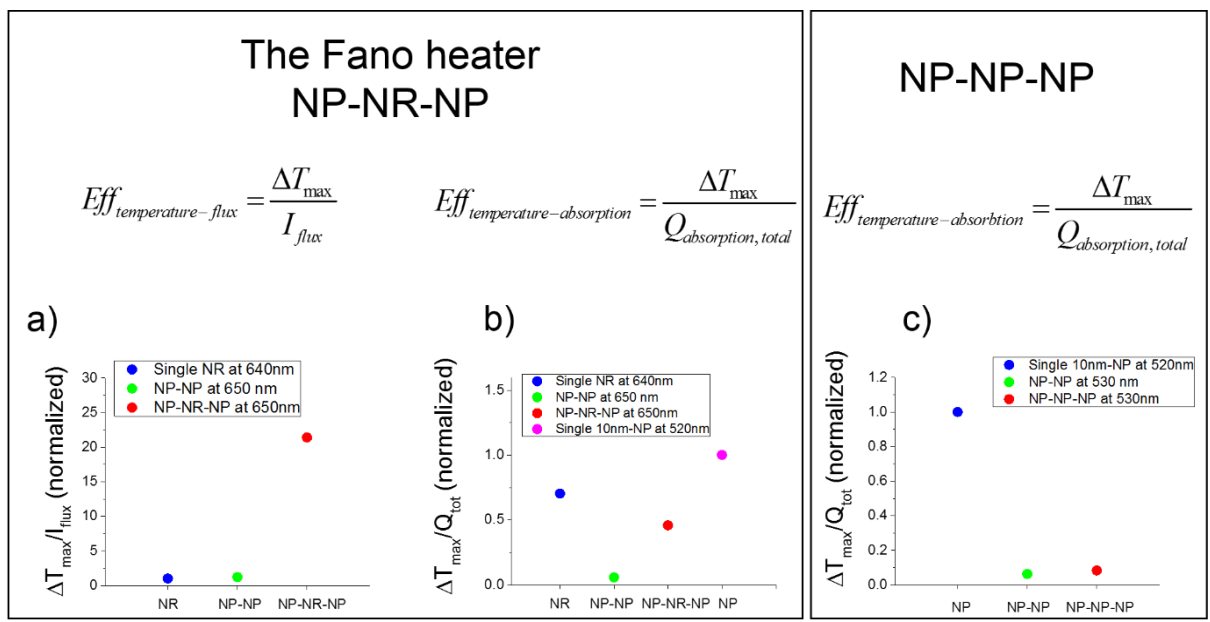

**Figure 13:** Power figures of merit for the NP-NR-NP trimer and for the previous NP-NP-NP complex. We see that the Fano complex has clearly an advantage over the NP-NP-NP structure.

## 6. Solvent-dependent photo-temperature

We note in this short section that the photo-generated increase of temperature and some of the figures of merit strongly depend on the matrix material. The above calculations were performed for water which is a very common matrix. However, experiments can be also carried out in organic solvents, which have typically lower heat conductivity. We give a few examples of thermal conductivities of the common solvents as Table S1 in Supporting Information. The advantage of organic solvents is that they will create stronger thermal isolation. Therefore, an optically-excited plasmonic system with an organic solvent can generate higher photo-temperatures. The simple scaling law for the maximum temperature in the CW-regime is

$$\Delta T_{\max} \propto \frac{1}{k_{t,matrix}}. \qquad (10)$$

We can see it easily for the case of a single NP (7). For a complex plasmonic system, the scaling law (10) is valid if $k_{t,matrix} \ll k_{t,Au}$. This condition is well satisfied for our plasmonic systems. Therefore, the local temperature distribution in a plasmonic system with an arbitrary matrix can be calculated from the above numerical results simply by multiplying by the scaling ratio $k_{t,w}/k_{t,matrix}$. For the common organic solvents given in Table S1, the maximum temperature in a system and the three figures of merit ($\alpha_{temperature\,gradient}$, $Eff_{temperature-flux}$ and $Eff_{temperature-absorption}$) will be increased by the factors $k_{t,w}/k_{t,matrix} \sim 3-6$. Simultaneously, the first figure of merit, $\alpha_{localization\,length}$, will not be strongly affected by the matrix.

## 7. Conclusions

We have investigated the possibility to create thermal hot spots by using electromagnetic hot spots in plasmonic nanostructures. For this, we place a small heater nanocrystal between two large antenna NPs. The formation of the thermal hot spots occurs under the peculiar conditions. In particular, the thermal hot spots appear in the trimer NP complex with small gaps in the NP-NP antenna due to the strong electromagnetic focusing. Another interesting possibility to form the thermal hot spots is the use of small nanorods as heater elements. In this case, we propose the NP-NR-NP complex operating in the regime of the plasmonic Fano effect. This Fano complex demonstrates strongly-enhanced figures of merit for localization of temperature.

ACKNOWLEDGMENTS

This work was funded by U.S. Army Research Office under Grant W911NF-12-1-0407, by Volkswagen Foundation (Germany) and via Chang Jiang Chair Professorship (China). L.K.K. was supported by the CMSS Fellowship Award at Ohio University.

# Supporting Information

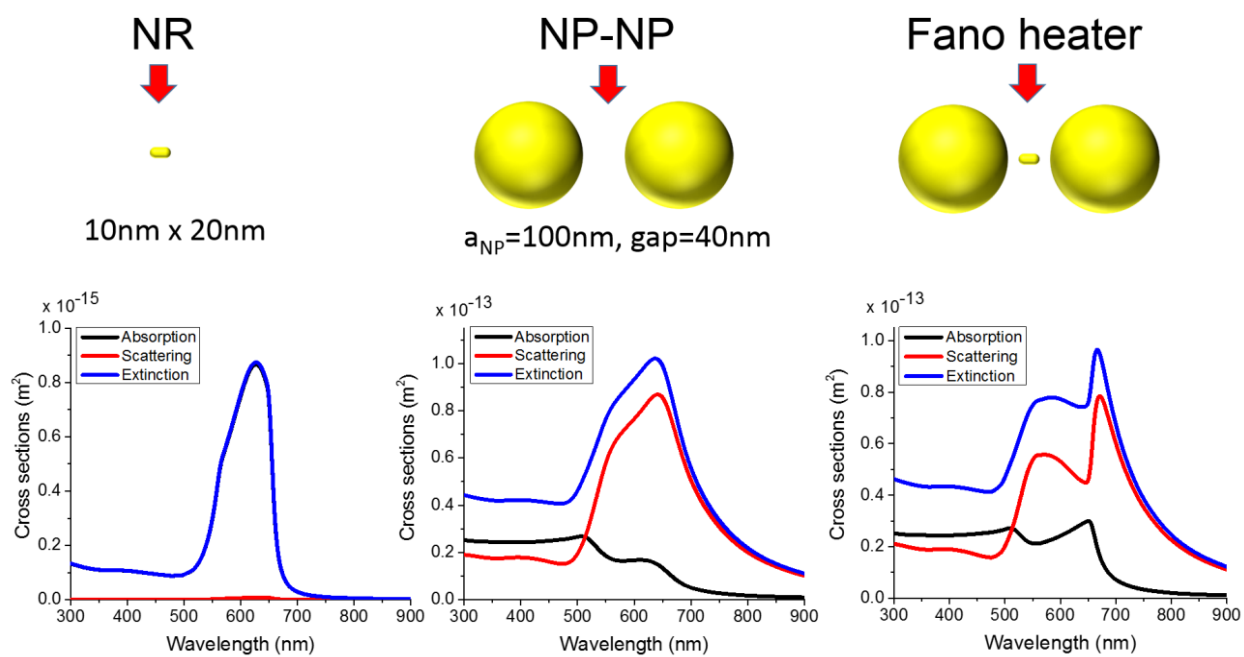

**Figure S1:** Calculated extinction cross sections for the three systems: NR, NP-NP and NP-NR-NP.

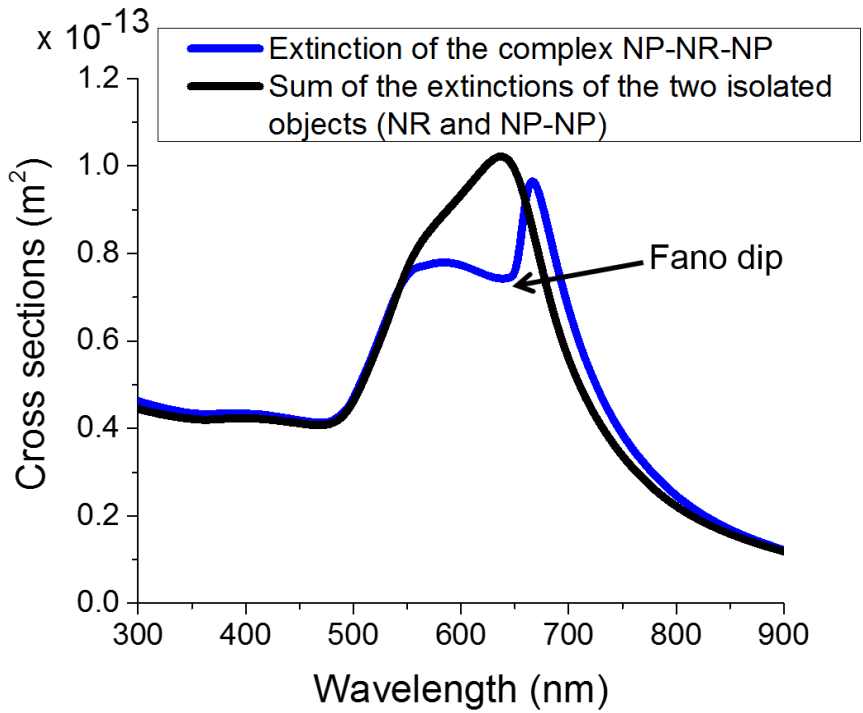

**Figure S2:** Comparison between the extinction cross section of the NP-NR-NP complex and the sum of the extinctions of the two isolated plasmonic systems (NR and NP-NP). We see the following: (1) The appearance of the Fano dip in the case of the NP-NR-NP complex; (2) The extinctions of the components are not additive, $\sigma_{NP-NR-NP} \neq \sigma_{NR} + \sigma_{NP-NP}$.

**Table S1:** Thermal conductivities of a few common solvents [S1].

| solvent | $k_t$ (W/mK) |
|---|---|
| water | 0.598 |
| methanol | 0.200 |
| ethanol | 0.169 |
| acetone | 0.161 |
| tetrachloromethane | 0.099 |

[S1] Bialkowski, S.E. *Photothermal spectroscopy methods for chemical analysis*, Volume 134 in *Chemical Analysis: A Series of Monographs on Analytical Chemistry and Its Applications,* J. D. Winefordner, Series Editor; John Wiley & Sons: Chichester, 1996.